\documentstyle[12pt]{article}
\input epsf
\def\plotone#1{\centering \leavevmode
\epsfxsize= 0.5\columnwidth \epsfbox{#1}}
\def\plottwo#1{\centering \leavevmode
\epsfxsize= 0.45\columnwidth \epsfbox{#1}}

\def\be{\begin{equation}}
\def\ee{\end{equation}}
\def\bea{\begin{eqnarray}}
\def\eea{\end{eqnarray}}

\def\cmm2{{\,\rm cm^{-2}}}
\def\cm2{{\,{\rm cm}^2}}
\def\cmm3{{\,{\rm cm}^{-3}}}
\def\gcmm3{{\,{\rm g\,cm^{-3}}}}

\def\muk{~\mu{\rm K}}

\def\ga{\mathrel{\mathpalette\fun >}}
\def\fun#1#2{\lower3.6pt\vbox{\baselineskip0pt\lineskip.9pt
  \ialign{$\mathsurround=0pt#1\hfil##\hfil$\crcr#2\crcr\sim\crcr}}}

\begin{document}
\baselineskip=12pt
\pagestyle{empty}
\begin{center}
\bigskip
\bigskip
\rightline{CITA-96-10}

\vspace{.2in}
{\Large \bf Cosmic Microwave Background Anisotropy Observing Strategy
Assessment}\\

\vspace{.2in}
Lloyd Knox\\
\vspace{.2in}

{\it Canadian Institute for Theoretical Astrophysics\\
University of Toronto, Toronto, Ontario, CANADA~~M5S 3H8}\\

\end{center}

\vspace{.3in}

\centerline{\bf ABSTRACT}
\medskip

\noindent
I develop a method for assessing the ability of an instrument, coupled
with an observing strategy, to measure the angular power spectrum of
the cosmic microwave background (CMB).  It allows for
efficient calculation of expected parameter uncertainties.  Related to
this method is a means of graphically presenting, via the ``eigenmode
window function'', the sensitivity of an observation to different
regions of the spectrum, which is a generalization of the traditional
practice of presenting the trace of the window function.  I apply these
techniques to a balloon-borne bolometric instrument to be flown this
spring (MSAM2).  I find that a smoothly scanning secondary is better
than a chopping one and that, in this case, a very simple
analytic formula provides a good (40\% or better) 
approximation to expected power spectrum uncertainties.

\newpage
\pagestyle{plain}
\setcounter{page}{1}

\section{Introduction}

Precision measurement of the angular power spectrum of the cosmic microwave
background (CMB) promises enormous scientific returns.  The expected results
for the next generation of satellite experiments have been studied extensively
\cite{mythesis, jungman}.  If one assumes that structure was formed
by\\
 gravitational instability then MAP \cite{map} and COBRAS/SAMBA
\cite{esa} should both measure $\Omega$, the Hubble constant and other
cosmological parameters to better than a few per cent.

These large sky coverage, map-making observations lend themselves to
easy analytic evaluation.  However, there are very important
balloon-borne and ground-based observations to be done over the next
few years for which the determination of expected power spectrum
uncertainties is not as straightforward.  The method presented here is
in principal analytic but the necessary high-dimensional linear
algebra requires numerical work in practice.  

One influence on the ability of any instrument to measure the
microwave background is its ability to separate the CMB from confusing
astrophysical foregrounds.  Much work has been done along these lines
\cite{forsep, dodelson}.  In contrast, relatively little attention has
been paid to the choice of {\it spatial} observing strategy and its
effect on expected power spectrum uncertainty.  The benefits of high
angular resolution have been emphasized \cite{mythesis} and it is
well-known that a rough guide to optimal sky coverage in a fixed time
is that which gives a signal-to-noise ratio per independent pixel of
unity.  However, the pixels are not independent, necessitating a more
sophisticated analysis.

One reason for this discrepancy between attention paid to frequency\\
strategy versus spatial strategy is that while the spectrum
of the CMB and confusing foregrounds have been known for a long time
(at least roughly), it is only recently we have gained some knowledge
of the angular power spectrum across a large range of angular scales.
Prior to any detection of anisotropy, the only possible
guidance on spatial strategy was that the optimal number
of points to observe is thirteen \cite{13lore}.
Now that our knowledge has improved substantially, it should 
be used for the choice of observing strategies.  

In section two I present the method and discuss the virtues of the
eigenmode window function.  In section three I derive a simple
analytic approximation for power spectrum uncertainty.  In section
four I apply the method to the case of MSAM2 as an example.  I use the
example to make a quantitative argument for a smoothly scanning
secondary as opposed to a chopping secondary, to present the eigenmode
window function, and to compare the calculated power spectrum and
cosmological parameter uncertainties with those from a simple analytic
formula.  An appendix details the calculation of the theory and noise
covariance matrices and also suggests a useful prescription for
normalizing window functions.

\section{Semi-analytic Method}

The data set can be modeled as consisting of signal and noise, $D_i =
S_i + N_i$.  Its covariance is:

\bea
 \langle D_i D_j \rangle & = & \langle S_i S_j \rangle +\langle N_i N_j \rangle
\\
  & = & C^S_{ij}+C^N_{ij}
\eea
For the simple case where the $D_i$ are measurements of the
temperature in directions enumerated by $i$, the signal covariance
matrix is only a function of the angle separating $i$ and $j$,
$\theta_{ij}$:
\be C^S_{ij} = C(\theta_{ij}) = \sum_l {2l+1 \over
4\pi} C_l P_l(\cos \theta_{ij})
\ee
where $C_l$ is called the angular power spectrum.  For reasons to be
discussed below, the data are generally not direct measurements of the
temperature in a given direction, but are linear combinations of
temperatures in several directions.  Calculation of $C^S$ and $C^N$ in
the general case is discussed in the appendix.

In general, the signal covariance matrix depends only on the observing
strategy and the angular power spectrum $C_l$.  We can think of the
angular-power spectrum as a function of parameters, $a_p$, either
physical ($\Omega_0, n_S, h, {\rm etc.}$) or phenomenological.  The
task at hand is to estimate, for a given experiment, the uncertainties
to expect on the parameters $a_p$.

Our task is simplified by transforming the data so that both the
signal and noise covariance matrices are diagonal.  The desired
transformation is the Karhunen-Loeve \cite{KL} transformation or,
as it has become known in CMB phenomenology, the signal-to-noise
eigenmode transformation \cite{bond}.
Bond \cite{bondsn} and White and Bunn
\cite{whitebunn} used it in their analyses of the COBE DMR maps.
Vogeley and Szalay have shown how it is useful for analysis of galaxy
redshift surveys \cite{sngal}.
The transformation is a non-unitary mapping of the data in the pixel
basis into the ``signal-to-noise eigenmode'' basis.
The transformation to $D_i'$ (the data in the S/N basis) is as follows:

\be
D_i' = R_{ij} ((C^N)^{-1/2})_{jk} D_k
\ee
where $C^N_{ij} \equiv \langle N_i N_j \rangle$,
R is the rotation matrix that diagonalizes \\
$M \equiv (C^N)^{-1/2} C^S (C^N)^{-1/2}$
and $C^S_{ij} \equiv \langle S_i S_j \rangle$.
It is straightforward to show that

\be
\langle D_i' D_j' \rangle = (\lambda_i + 1)\delta_{ij}
\ee
where $\lambda_i$ are the eigenvalues of the S/N matrix, $M$.
Thus we have the desired transformation, since the signal and noise
covariance matrices are now diagonal.

We can now think of the experiments as measuring the $\lambda_i$ which
are a function of the power-spectrum and therefore of the parameters,
$a_p$.  We can build a quadratic estimator for $\lambda_i$ from $D_i'$
by squaring it and subtracting unity:

\be
\lambda_i^{\rm est} = D_i'^2 - 1
\ee
This estimator has variance $2(\lambda_i+1)^2$ since $D_i'$ is a Gaussian
random variable with variance $\lambda_i+1$.

 From here it is straightforward to calculate uncertainties on the parameters.
First calculate the curvature matrix $\alpha$:

\be
\alpha_{pp'} = \sum_i {1\over \sigma^2(\lambda_i)} {\partial \lambda_i \over
\partial a_p} {\partial \lambda_i \over \partial a_{p'}}.
\label{eqn:alpha}
\ee
where $\sigma^2(\lambda_i) = 2(\lambda_i+1)^2$.
Then invert to get the desired result, the
parameter covariance matrix \cite{numrep,jungman,T96}:

\be
C^P_{pp'} =  \left(\alpha^{-1}\right)_{pp'}.
\label{eqn:alpha2cov}
\ee
Equation \ref{eqn:alpha2cov} is strictly true only if our estimates of
$a_p$ are Gaussian random variables which will be the case if the
eigenvalues depend linearly on $a_p$.  In general the $\lambda_i$ 
have non-linear dependences on physical parameters, but if the
data constrain the parameters sufficiently then
the non-linear effects will be unimportant.

The covariance matrix is simply given by the inverse of the curvature
matrix for the case when we have no prior information on the
parameters.  At the other extreme, if the parameters other than $a_p$
are perfectly known then $\sigma^2(a_p) = 1/\alpha_{pp}$.  The
intermediate case is easily treated if the prior information is
expressed as a covariance matrix $C_{\rm prior}$.  Then

\be
\sigma^2(a_p)=\left(\left(\alpha+C_{\rm prior}^{-1}\right)^{-1}\right)_{pp}.
\ee

To summarize, the procedure is straightforward.  First choose a\\
parametrized theory and calculate the signal and noise covariance
matrices.  This step requires a specific choice of theory parameters.
Calculate the rotation matrix, $R$, the eigenvalue spectrum,
$\lambda_i$, and its derivatives.  Calculate the curvature matrix, add
any prior information and invert to get the estimated parameter
covariance matrix.

The imaginary analysis of the data that I used above to derive
Eq. \ref{eqn:alpha} neglected off-diagonal terms that would be
included in a full likelihood\\ analysis.  A Taylor series expansion in
the log of the likelihood function, about its maximum, 
leads to the result \cite{KLrev} 

\be
2 \alpha_{pp'} = {\rm tr}[(C^S+C^N)^{-1}{\partial C^S \over \partial a_p}
(C^S+C^N)^{-1}{\partial C^S \over \partial a_{p'}}]
\label{eqn:trace}
\ee
where tr stands for the trace.
With a little algebra this becomes:
\be
\alpha_{pp'} = \sum_{i,j} {({\partial \lambda \over \partial a_p})_{ij}
({\partial \lambda \over \partial a_{p'} })_{ji} \over
                          2(\lambda_i+1)(\lambda_j+1)}.
\label{eqn:2alpha}
\ee

The difference between the two results (Eq. \ref{eqn:alpha2cov} and
Eq.  \ref{eqn:2alpha}) is simple to understand.  The imaginary
analysis above neglected the off-diagonal terms that arise when we
move in parameter space, off of the parameters for which the signal
matrix is diagonalized.  In other words, although $\lambda$ is
diagonal by design, its derivatives are not.  Since the maximum
likelihood is the ``best'' \cite{endnote} estimator, the parameter
variances following from Eq. \ref{eqn:2alpha} will be smaller
than those from Eq. \ref{eqn:alpha2cov}.   I will refer to
the use of Eq. \ref{eqn:alpha2cov} as the diagonal approximation
and to Eq. \ref{eqn:2alpha} as ``exact'', with quotations due
to the implicit approximation of linear parameter dependence
already mentioned.

Eq. \ref{eqn:2alpha} is an improvement on Eq. \ref{eqn:trace} because
of the compression it allows;  the modes below some level
of signal-to-noise can be discarded.  Because it neglects the
off-diagonal terms, the diagonal approximation is even more convenient
and in the applications that follow I show that the information
loss from neglecting the off-diagonal terms is small.  

Another advantage of the accuracy of the diagonal approximation is
that it allows for easy visualization of how the observations will
probe the spectrum.  The window function $W_{lij}$ relates the power
spectrum and its derivatives to the signal covariance matrix and its
derivatives:

\be
{\partial C^S \over \partial a_p}=
\sum_l{\partial C_l \over \partial a_p}{2l+1 \over 4\pi}W_{l}
\label{eqn:pixbasis}
\ee

In the pixel basis the window function is complicated.  The usual
practice for indicating what region of the spectrum is probed is to
plot only its trace -- a procedure which neglects possibly important
off-diagonal terms.  It is possible to do much better in the eigenmode
basis.  From Eq. \ref{eqn:pixbasis} follows:

\be
{\partial \lambda_{ij} \over \partial a_p}=
\sum_l {\partial C_l \over \partial a_p} W_{lij}' {2l+1\over 4\pi}
\label{eqn:eigenbasis}
\ee
where the prime indicates the eigenmode basis.
In this basis one can safely ignore the off-diagonal elements,
reducing the number of relevant dimensions to two (one mulitpole
moment index plus one pixel index), allowing for\\ visualization of all
the important parts of the window function.

\section{Simple Analytic Methods}

For a map with uniform full-sky coverage and Gaussian
angular resolution, $\sigma_b$, the eigenmodes are the
spherical harmonics with eigenvalues equal to
$w C_l e^{-l^2\sigma_b^2}$ where $w$ is
the weight per solid angle.  There are $2l+1$ modes
for each $l$ and thus $\sigma^2(\lambda_i)
= 2(\lambda_i+1)^2$ leads to the formula (derived by other means in
\cite{mythesis}):

\be
\sigma^2(C_l) = {2\over 2l+1}\left(C_l + w^{-1} e^{l^2\sigma_b^2}\right)^2.
\label{eqn:analytic}
\ee

The effect of observing only a fraction of the sky, $f$, can be
approximately described by increasing the variance of $C_l$ by
$f^{-1}$ since the number of modes is roughly proportional to the area
\cite{jungman, hobson, smoot}.  Jungman {\it et al.}
\cite{jungman} used this formula and the analogs of 
Eq. \ref{eqn:alpha} and \ref{eqn:alpha2cov} to 
calculate standard errors for an
eleven parameter adiabatic gravitational instability model.  The
partial-sky corrected version of Eq. \ref{eqn:analytic}
must be used with care, however.  The modes are no
longer spherical harmonics and therefore any estimate for $C_l$ will
be correlated with that of $C_l'$ -- a correlation with range $\Delta
l \approx 2\pi/\theta$ where $\theta$ is a linear dimension of the
observed field \cite{T95}.  The equation only makes sense when the
spectrum is binned with $\Delta l \ga 2\pi/\theta$.  

\section{Application to MSAM2}

In this section I apply the above formalism to the particular case of
the second Medium Scale Anisotropy Measurement instrument (MSAM2).
The MSAM instruments are balloon-borne off-axis Cassegraine telescopes
with bolometric radiometers.  The MSAM1 instrument is described in
\cite{ms1inst}.  Detections have been reported from two of the flights
\cite{ms1det} with the second flight confirming the results of the
first \cite{ms1conf}.  The MSAM2 instrument uses the same gondola as
MSAM1 but has a different radiometer and secondary mirror.

The largest component of atmospheric contamination depends only on
elevation and is slowly varying in time.  Thus a standard technique to
reduce atmospheric contamination is to rapidly sample a stretch of sky
at constant elevation by motion of a secondary mirror from $-\theta_c$
to $\theta_c$.  Only linear combinations of the data that have no
sensitivity to a spatially homogeneous signal are kept for further
processing.  In some cases those combinations sensitive to a gradient
are discarded as well.  Each linear combination corresponds to a
``synthesized antenna pattern''.  The MSAM1 secondary motion was a
three point chop.  From the time stream, two antenna patterns were
synthesized.

The MSAM2 secondary motion will be a triangle wave pattern of
period $T$.  Here we model the time stream of data as follows:

\be
d(t) = A_0(\theta_i)/2 + \sum_{\mu=1}^8 D^c_\mu(\theta_i)
\cos(2\pi \mu {t\over T}) +
\sum_{\nu=1}^8 D^s_\nu(\theta_i) \sin(2\pi \nu {t\over T}).
\label{eqn:tstream}
\ee
where $\theta_i$ is a slowly changing function of time and
refers to the point on the sky observed when the secondary
is in its central position.

We cut the Fourier decomposition off at $\nu=\mu=8$ because the peak-to-peak
chop amplitude is 8 beam full-widths; higher frequency modes would
have very little signal.  Since the secondary
motion is symmetric about $t = T/2$, the asymmetric Fourier modes
will contribute zero signal.  Thus the odd $\nu$ components are
ignored.

If we assume that the noise in the time stream $d(t)$ is white, then
the noise in each of the above modes will be independent; the noise
covariance matrix $C^N$ will be diagonal (see appendix).  A better
model would also have terms that vary in time but not in space, as is
the case for instrumental and atmospheric drifts.  Having to fit for
the coefficients of these terms would induce correlations in the noise
covariance matrix $C^N$.  Given a model of the drifts it is possible
to calculate $C^N$ but here I ignore these
effects and take the matrix to be diagonal.  From a model of
the bolometer and the foregrounds \cite{dodelson} (a one component dust
model) we expect the sensitivity to CMB to be NET\footnote{The
standard error of the observed temperature is given by ${\rm
NET}/\sqrt{t_{\rm obs}}$ where $t_{\rm obs}$ is the observing time.}
$\simeq 266 \muk \sqrt{\rm sec}$.

Although the noise matrix is diagonal, the signal matrix is not.  The
off-diagonal correlations exist for three conceptually distinct
reasons.  First, the Fourier decomposition in Eq. \ref{eqn:tstream} is
for functions of period $T$ but because of the triangle wave motion of
the secondary, the same stretch of sky is scanned twice in that one
period.  The $D_\mu^c(\theta_i)$ and $D_\mu^s(\theta_i)$ are the
cosine and sine coefficients of the sky sampled from $\theta_i -
\theta_c$ to $\theta_i + \theta_c$, with fundamental frequency half
what it would be for a Fourier decomposition.  Second, because the
central position of the chopper changes over time, the
decompositions do not all share the same origin.  Thus, even if
the fundamental mode had the right frequency, the modes with $\theta_i
\ne \theta_j$ are not orthogonal.  Third, since the sky is not
``white'' like the noise (there are intrinsic correlations) the
different modes are correlated -- even when $|\theta_i - \theta_j| >>
2\theta_c$.  The calculation of the signal covariance matrix, $C^S$,
is described in the appendix.

\subsection{Optimal Motion of the Secondary}

\begin{figure}[bthp]
\plotone{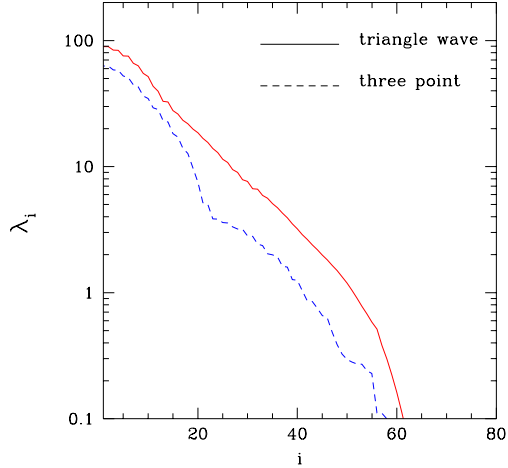}
\caption[S/N Eigenvalue Spectra] {\baselineskip=10pt 
The S/N eigenvalue spectrum for
a three-position secondary and for a smoothly scanning secondary.}
\label{fig:eval}
\end{figure}

Is it better to move the secondary in a step motion between two or
three spots, or to smoothly scan it back and forth?  Here I have
addressed that question by computing the S/N eigenvalue spectra, shown
in Figure \ref{fig:eval}, for a three-point chop and a triangle wave.

The two curves shown in Fig. 1 are both S/N spectra for observing
strategies that are the same in all respects except for the motion of
their secondaries.  They both point at a declination of $\delta = 78$
degrees on the transit meridian as the secondary moves back and forth
$\pm 80'$ at constant declination (an approximation to the actual
motion in cross-elevation).  The beam is taken to be a
Gaussian with $\sigma_b= 20'/(\sqrt{8ln2})$.  The sky is observed in
this manner for five hours.  The rotation of the sky leads to coverage
of a strip 15.6 degrees long.  For a theory I took a flat spectrum
with:

\be
\delta T_l^2 \equiv {l(l+1)C_l \over 2\pi} = (50 \muk)^2.
\ee

The dashed curve is for the secondary that executes a three-point chop.
The single difference and double difference signals are analyzed.  The
solid curve is for the triangle wave motion.  In this case the signals
from the 12 different synthesized antenna patterns were analyzed.  One
can see that the chopping secondary is inferior to the smoothly
scanning ones since its eigenvalue spectrum is lower for every mode.

As a rough guide to the power-spectrum sensitivity of the experiment,
we can simply count the number of modes with $\lambda_i > 1$.  This is
because modes with $\lambda_i >>1$ all have the same fractional error
($\sqrt{2}$) while those with $\lambda_i << 1$ have very large
uncertainties.  For a more precise comparison we can perform the sum

\be
\sigma^2(A) = \left( \sum_i {\lambda_i^2 \over 2(\lambda_i+1)^2}\right)
^{-1}
\label{eqn:amp}
\ee
where $A$ is the amplitude of the spectrum whose shape has been assumed.
For the triangle wave spectrum $\sigma(A) = 0.22$.

It is easy to see how sensitivity to the amplitude of the spectrum will
change with varying sky coverage.  Increasing the sky coverage by a
factor of $n$ will increase the number of modes by a factor of $n$.
If the observing time remains fixed, then the noise in each pixel will
increase, reducing each eigenvalue by a factor of $n$.  From Fig. 1,
we see that it would be highly beneficial to greatly increase the sky
coverage.  

The uncertainty in the amplitude has a very shallow minimum at $n =
51$ of $\sigma(A) = 0.087$.  With such large sky coverage, the noise
in each beam-size pixel is $97 \muk$.  This is much larger than
the signal in any of the synthesized antenna patterns but is just
slightly smaller than the ``stare mode'' (undifferenced) signal of
$120 \muk$.  Thus we see that if one is solely interested in
measuring an amplitude, setting the {\it undifferenced} signal-to-noise
ratio to unity gives a near optimal sky coverage.  However, the
need to understand the inevitable non-idealities of the data
argues against spreading the weight this thinly.

\subsection{Band Estimates}

\begin{figure}[bthp]
\plotone{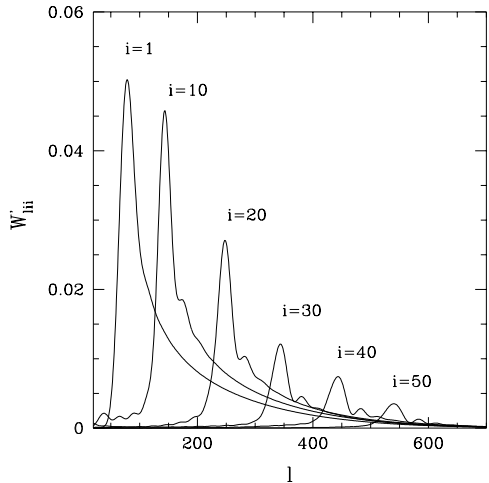}
\caption[Band Uncertainties] {\baselineskip 10pt
A few modes of the eigenmode window function}
\label{fig:win}
\end{figure}

The eigenvalue spectrum shows the sensitivity of the experiment 
to the overall amplitude of the spectrum but gives no indication 
of sensitivity to shape.  To understand what regions of the spectrum
are being probed it is useful to look at the eigenmode window
function, shown in Fig. \ref{fig:win} for the triangle
wave observations of the previous subsection.  As expected, 
the peaks move to higher values of $l$ as the mode number increases.

\begin{figure}[bthp]
\plottwo{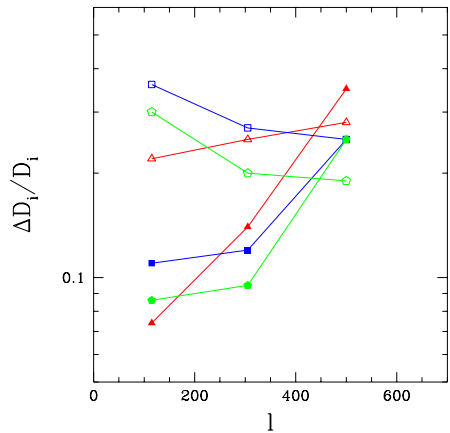}
\caption[Band Uncertainties] {\baselineskip=10pt
The expected fractional standard error for the three bands for the
``parallel'' (triangles) and ``perpendicular'' (squares) observing
strategies (see text).  The pentagons are the corresponding results
from the analytic approximation.  Filled symbols are for sixteen times as
much area as for the open symbols; observing time is fixed at 5
hours.}
\label{figeqal}
\end{figure}

Traditionally the region probed by an experiment has been indicated by
the diagonal elements of the pixel basis window function, 
$W_{lii}$ \cite{windows}.  This is
adequate if all that is important is the $rms$ of the data.  However,
off-diagonal correlations are important as well and can significantly
alter what regions of the spectrum are probed \cite{srednicki}.  In
particular it should be noted that as signal-to-noise ratio increases
more information starts coming from the smaller scales -- an effect
that should be clear from Fig.  \ref{fig:win}.  To quantify how well
the experiment is probing different parts of the spectrum it is useful
to estimate the uncertainty in several bands.  To simplify
interpretation it is useful to choose the bands with widths greater
than the eigenmode window function widths since that will reduce the 
correlations between estimates.  For Fig. \ref{figeqal} we have 
parametrized the spectrum as

\be
\delta T_l = \cases{D_1 &if $20 \le l < 210$\cr
                   D_2 &if $210 \le l < 400$\cr
                   D_3 &if $400 \le l < 810$\cr}
\ee

\noindent and taken $D_1 = D_2 = D_3 = 50 \muk$.
Note that estimating the power in a given band is very different from
estimating the ``band-power'' for a synthesized antenna pattern.  In
the former we are using all the data to constrain part of the
spectrum.  In the latter we are using part of the data to constrain
the entire spectrum.

The open triangles in Fig.~\ref{figeqal} show the expected 
uncertainties for the triangle wave observing strategy described in
the previous subsection.  I will refer to this strategy as
``parallel'' since the sky coverage is swept out parallel to the
motion of the secondary.  The open squares are for a strategy where
the sky is swept out perpendicular to the secondary motion.  Each case
has the same area: 5.2 sq. degrees, the same sensitivity, observing
time, beam size, and even the same synthesized antenna patterns.
The one long dimension of the ``parallel'' strategy is the reason
for its superiority in the lowest spatial frequency bin.

The band estimates allow us to see the effect of varying sky coverage
in more detail.  The closed symbols in Fig. \ref{figeqal} are the
expected uncertainties for observing 16 patches of the sky in the same
way, with the same total observing time.  The 16 patches are assumed
to be far enough apart that the correlations between them can be
ignored.  In this case, the calculation is a simple extension of that
done for the open symbols: the number of modes is increased by 16 and
the S/N eigenvalues are all reduced by a factor of 16.  Note that as
expected, the lower the spatial frequency of the band, the more it
benefits from extra sky coverage.

The ``parallel'' and ``perpendicular'' strategies are two extremely
different ways of uniformly covering 5.2 sq. degrees.  Despite this
fact, the analytic approximation (pentagons), which does not depend on
how the sky is covered, is good for both of them to better than 40\%.
Where the strategies are most different ($D_1$), it splits the
difference!  The results from the diagonal approximation 
(not shown) differ by at most 10\%.  

\subsection{Cosmological Parameters}

The sky coverage planned for the MSAM2 flight is too small to allow
for simultaneous determination of a large number of cosmological
parameters.  However, it is interesting to vary a limited set of
parameters and contrast the results from the ``exact'' formula with
the diagonal approximation and the analytic approximation.
To calculate the standard errors in the Table I
varied the quadrupole, $\langle Q\rangle$, the scalar power-law index
$n_S$ and $\Omega_{\rm tot} = \Omega_0+\Omega_{\Lambda}$ around their
standard CDM values of $\langle Q \rangle = 20 \muk$, $n_S = 1$ and
$\Omega_{\rm tot} = 1$ \cite{arthur}.  
For prior information, I assumed that COBE-DMR
determined the quadrupole to within 20\% \cite{bennett96}.  
Calculations were done for nominal and 16 times sky coverage.

\begin{table}[htb]
\centering
\caption{Constraints on Cosmological Parameters}\label{tab:cosmoparam}
\vspace{1ex}
\begin{tabular}{ccccc} \hline \hline
\multicolumn{1}{c}{Sky Coverage} &
\multicolumn{1}{c}{Method} &
\multicolumn{1}{c}{$ \Delta \langle Q \rangle$} &
\multicolumn{1}{c}{$\Delta n_S$} &
\multicolumn{1}{c}{$\Delta \Omega_{\rm tot}$} \\ \hline
1  & Diagonal  & 4.0 & 0.27  & 0.35 \\
1  & ``Exact'' & 4.0 & 0.22  & 0.28 \\
1  & Analytic  & 4.0 & 0.18  & 0.22 \\ \hline
16 & Diagonal  & 4.0 & 0.15  & 0.17 \\ 
16 & ``Exact'' & 3.9 & 0.12  & 0.14 \\
16 & Analytic  & 3.9 & 0.11  & 0.12 \\
\hline
\end{tabular}
\end{table}

It is not surprising that the diagonal approximation does worse here
than for the band estimates since the diagonal is equal to the
``exact'' when the parameter is a normalizing constant.  But even with
these cosmological\\ parameters it is still good to 25\%, bolstering the
claim that the diagonal elements of the eigenmode window function hold
most of the information.  The analytic approximation, off by 20\% at
the worst, is even better here than for the band estimates.  Note that
the constraint on $\langle Q \rangle$ comes almost entirely from the
prior information.

\section{Discussion}

The simple analytic approximation worked quite well for the cases
studied.  The approximation will be worse for observations where the
differencing throws away more information, the off-diagonal noise is
more important, or the sky coverage is less uniform.  Differencing
will throw away more information when the secondary throw to beam fwhm
ratio is smaller or, as was explicitly shown, if the secondary chops
discontinuously.  On the other hand, we can expect the analytic
formula to be a very good approximation for\\ partial sky map-making
observations, such as those to come from Long Duration Ballooning.

Besides a method for estimating parameter uncertainties I have also\\
presented a useful tool for understanding how an observation probes
the spectrum.  Combined with a graph of the eigenvalue spectrum, the
diagonal elements of the eigenmode window function allow one to
immediately see the $\ell$-space resolution, the sensitivy to
different regions of the spectrum, and the dependence of that
sensitivity to varying sky coverage.  The traditional use of the
diagonal elements of the pixel basis window function can be misleading if the
experiment has a high signal-to-noise ratio.

The bands in Fig. \ref{figeqal} were chosen with large widths to reduce
correlations.  However, these large widths mean some resolution has
been discarded and thus this is not a good solution to the problem of 
presenting data in a compact, yet complete, manner.  Further work
along these lines is needed.

\bigskip\bigskip\bigskip
\section{Acknowledgments}
\addcontentsline{toc}{section}{Acknowledgments}

I would like to thank Andrew Jaffe, Max Tegmark, Martin White and my
collaborators on MSAM2 for useful discussions as well as Grant Wilson,
Shaul Hanany and Adrian Lee whose questions stimulated this work.

\appendix

\section{Covariance Matrix Calculation}

Given an observing strategy and a theory, we can construct the
theoretical correlation matrix.  Here I show how that calculation is
done in general.  For simplicity of notation, we assume that the
telescope is pointed at $\theta_i$ for only one cycle of the secondary
before moving on to $\theta_j$.  In reality, many cycles are completed
before the telescope's pointing has changed significantly.  Assume
that the detectors are sampled $n$ times in one period $T$ and
$\theta_{ia}$ is the direction on the sky observed on the $a^{\rm th}$
sample of the $i^{\rm th}$ secondary cycle.  The multiple antenna
patterns are synthesized by weighting the $n$ samples of each cycle by
the weight vector $w_{\mu a}$:

\be
D_{\mu i} = {1\over n}\sum_{a=1}^n d_{ia} w_{\mu a}.
\ee
For example, the weight vectors for $D^c_\mu(\theta_i)$ in
Eq. \ref{eqn:tstream} are

\be
w_{\mu a} = \cos\left( 2\pi \mu a/n \right)
\ee

To avoid proliferation of indices, we will now write, e.g., $D_{\mu
i}$ as $D_i$ where $i$ is now understood to run over just pixels, just
antenna patterns, or both, according to context.  The data is once
again modelled as due to signal from the microwave background, $S_i$
and noise from the atmosphere and instrument $N_i$.  The time stream,
$d_{ia}$ is also split into signal $s(\theta_{ia})$ and noise
$n_{ia}$.

The two-point theoretical correlation function $C^S_{ij} \equiv
\langle S_i S_j \rangle$ can be easily related to the intrinsic
two-point correlation function $C^s(\theta_{ai},\theta_{bj}) \equiv
\langle s(\theta_{ai}) s(\theta_{bj})\rangle$:

\be \label{ceqn}
C^S_{ij} =
\left({1\over n}\right)^2\sum_a \sum_b
\langle s(\theta_{ia}) s(\theta_{jb})\rangle w_{\mu a} w_{\nu b}.
\ee

\noindent By isotropy in the mean, $C^s(\theta_{ai},\theta_{bj})$ only
depends on the angular distance, $\theta_{iajb}$,
between $\theta^i_a$ and $\theta^j_b$.
Thus, the correlation function can be decomposed
into Legendre polynomials and Eq. \ref{ceqn} can be
rewritten:

\be
C^S_{ij} = \sum_l {2l+1 \over 4\pi}C_l W_{lij}
\ee

\noindent where

\be
W_{lij} \equiv  \left({1\over n}\right)^2\sum_a \sum_b
P_l\left(\cos(\theta_{iajb})\right)w_{ia} w_{jb}.
\ee

If the noise is white (uncorrelated from sample to sample)
then the noise matrix

\be
\langle \langle n_i n_j \rangle = n { {\rm NET}^2 \over T} \delta_{ij}
\ee
and therefore

\be
C^N_{ij} \equiv \langle N_i N_j \rangle = { {\rm NET}^2 \over T}
\left({1\over n}\right)\sum_k w_k^2 \delta_{ij}.
\label{eqn:cn}
\ee

A convenient normalization prescription for the weight vector is to
set $\sum_i w_i^2/n$ to a constant because then the variance in the
noise is the same for every antenna pattern.  Taking that constant to
be unity, we find the variance in the noise for each antenna pattern
is simply ${\rm NET}^2/t$ where $t$ is the total observing time -- the
same formula as for ``stare mode''.  This weight vector normalization
is equivalent to a window function normalization and is actually a
very sensible one.  Since the NET is the same for all antenna
patterns, the window functions with the higher signal-to-noise ratios
will have larger amplitudes.  Thus plotting the window functions
normalized in this way allows for quick graphical comparson of both
the l-space coverage and relative sensitivities of the various antenna
patterns.


\begin{thebibliography}  {ucsc}

\bibitem{mythesis} L. Knox, {\it Phys. Rev. D} {\bf 52}, 4307 (1995).

\bibitem{jungman} G. Jungman, M. Kamionkowski, A. Kosowsky and D.N. Spergel,
{\it Phys. Rev. Lett.} {\bf 76}, 1007 (1996);  Jungman et al.,
astro-ph/9512139.

\bibitem{map} E.L. Wright, G. Hinshaw and C.L. Bennett, Ap. J. {\bf
458}, L53 (1996); http://map.gsfc.nasa.gov/.

\bibitem{esa} M. Bersanelli, 
{\it et al.}, ESA Report 
D/SCI(96)3;\\ 
http://astro.estec.esa.nl/SA-general/Projects/Cobras/cobras.html

\bibitem{forsep} W.N. Brandt, C.R. Lawrence, A.C.S. Readhead, J.N.
Pakianathan, and T.M. Fiola, Ap. J. {\bf 424}, 1 (1994); M. Tegmark
and G. Efstathiou, astro-ph/9507009, (1995).

\bibitem{dodelson} S. Dodelson, 
astro-ph/9512021.

\bibitem{13lore} N. Kaiser, CITA, personal communication.


\bibitem{KL} Karhunen, K. 1947, {\it Uber lineare Methoden in der
Wahrschleinlichkeitsrechnung}, Helsinki: Kirjapaino oy. sana

\bibitem{bond} J.R. Bond, {\it Cosmic Structure Formation and the
Background Radiation}, in ``Proc. IUCAA Dedication Ceremonies,''
Pune, India, Dec. 1992, ed. T. Padmanabhan, Wiley (1994).

\bibitem{bondsn} J.R. Bond,  Phys. Rev. Lett. {\bf 74} 4369 (1995).

\bibitem{whitebunn} M. White and E. Bunn, astro-ph 9503054.

\bibitem{sngal} M.S. Vogeley and A.S. Szalay, Ap. J., in press (1996).

\bibitem{numrep} W.H. Press, B. Flannery, S.A. Teukolsky and
W.T. Vetterling, ``Numerical Recipes in Fortran'', 2nd ed.

\bibitem{T96} M. Tegmark, astro-ph/9511148, to appear in Proc. Enrico
Fermi, Course CXXXII, Varenna, 1995.

\bibitem{KLrev} M. Tegmark, A.~N. Taylor and A.~F. Heavens,
astro-ph/9603021

\bibitem{endnote} The ML-estimator is asymptotically (in the limit of
large data sets) the linear unbiased estimator with the smallest
variance.  See the review \cite{KLrev} and references therein.

\bibitem{hobson} M.P. Hobson and Joao Magueijo,
astro-ph/9603064.

\bibitem{smoot} G.~F. Smoot,
astro-ph/9505139.

\bibitem{T95} For an extensive discussion of spatial frequency resolution
see M. Tegmark, {\it Mon. Not. Roy. Astr. Soc.} {\bf 280}, 299 (1996),
or \cite{hobson}.
 
\bibitem{ms1inst} D.~J. Fixsen, {\it et al.}, Ap. J. (1996), 
astro-ph/9512006.

\bibitem{ms1det} E.~S. Cheng, {\it et al.}, Ap. J. {\bf 422}, L37 (1994);
E.~S. Cheng, {\it et al.}, Ap. J. {\bf 456}, L71 (1996).

\bibitem{ms1conf} C.~A. Inman, {\it et al.}, 
astro-ph/9603017.

\bibitem{windows} J.R. Bond, {\it et al.}, Phys. Rev. Lett. {\bf 66}, 2179
(1991).

\bibitem{srednicki}  The importance of off-diagonal correlations
for the case of two-dimensional sky coverage by a chopping experiment
was emphasized by M. Srednicki, M. White,
D. Scott and E. Bunn, Phys. Rev. Lett. {\bf 71}, 3747 (1993).

\bibitem{bennett96} C.~L. Bennett, {\it et al.}, COBE Preprint 96-01,
astro-ph/9606017

\bibitem{arthur}  The angular power spectrum and its derivatives were
kindly supplied by the authors of \cite{jungman}. 


\end{thebibliography}
\end{document}